\documentstyle[prd,aps,psfig]{revtex}
\begin{document}
\input epsf
\twocolumn[\hsize\textwidth\columnwidth\hsize\csname
@twocolumnfalse\endcsname
\title{{\hfill \small hep-ph/0102264}\\
$~$\\ Nonthermal production of gravitinos and inflatinos}
\author{H.P.~Nilles$^{(1)}$, M.~Peloso$^{(1)}$, and L.~Sorbo$^{(2)}$}

\address{$^{(1)}${\it Physikalisches Institut, Universit{\"{a}}t Bonn
Nussallee 12,
D-53115 Bonn, Germany.}}
\address{$^{(2)}${\it SISSA/ISAS, via Beirut 2-4, I-34013 Trieste, Italy, 
and INFN, Sezione di Trieste, via Valerio 2, I-34127 Trieste, Italy.}}

\maketitle
\begin{abstract}
We explicitly calculate nonthermal gravitino production during the preheating period in the inflationary Universe. Contrary to earlier investigations, we consider a two--field model to separate the mechanisms of supersymmetry breaking and inflation. We show that the superpartner of the inflaton is significantly generated, while the gravitino production is considerably smaller. Nonthermal production of gravitinos seems thus less worrisome than recently claimed.
\end{abstract}
\vskip2pc]

\def\gsim{\;\raise0.3ex\hbox{$>$\kern-0.75em\raise-1.1ex\hbox{$\sim$}}\;}
\def\lsim{\;\raise0.3ex\hbox{$<$\kern-0.75em\raise-1.1ex\hbox{$\sim$}}\;}

\section{Introduction}

Inflationary models, when embedded in particle--physics motivated schemes,
are usually constrained by the requirements of consistency with the
phenomenology of the later evolution of the Universe. In particular, in
the context of supergravity theories the parameters of the inflationary
sector have to be restricted in order to avoid the thermal overproduction
of gravitinos at reheating. These constraints, in the case of models with
gravity--mediated supersymmetry breaking, can be summarized as an upper
limit on the reheating temperature of the order of $10^9$~GeV (for a
review, see~\cite{tm}). In the last years it was shown that, during the
stage of coherent oscillations of the inflaton condensate right after
inflation, preheating can lead to an efficient nonthermal production of
fermions~\cite{gprt}. In particular, it was recently
argued~\cite{kklv1,grt1,grt2,kklv2} that the nonthermal production of
$1/2$--helicity component of gravitinos can in some cases be much more
efficient than the thermal one, thus worsening the gravitino problem. The
investigation that led to the present letter was actually triggered by
this last statement. Explicit calculations of the amount of gravitinos
produced at preheating were performed so far only in models with one
superfield and without supersymmetry breaking in the vacuum. Therefore,
the conclusions about the production of longitudinal gravitinos in these
models could be misleading, since there is no longitudinal gravitino in
the vacuum of the theory. Thus, one might wonder whether preheating could
actually lead to a production of harmless inflatinos rather than of
dangerous gravitinos.

In order to discriminate between inflatino and gravitino production it is
necessary to consider more realistic schemes. The simplest possibility is
to consider two separate sectors, one of which drives inflation, while the
second is responsible for supersymmetry breaking today. In the present
analysis we study the situation in which the two sectors communicate only
gravitationally. The calculation of nonthermal gravitino production in
this scheme requires substantial work. First, one has to develop a new
formalism, to be able to clearly define and compute the production in
systems with several coupled fields. Subsequently, an extended numerical
investigation has to be carried out to obtain reliable results. In fact,
this procedure led us to conclusions that do not confirm earlier
analytical estimates~\cite{kim}.

The paper is structured as follows. First, we define the specific model
considered, analyzing the evolution of the scalar fields. Then, we discuss
how the definition of the gravitino varies in time according to the
evolution of the background. We finally compute the spectrum of the
gravitinos produced at preheating, concluding that what in the previous
works was believed to be gravitino overproduction should rather be
regarded as inflatino production.

\section{Gravitino production in a two field model}

The system we are considering has the matter content of two superfields $\Phi$ and $S$, with superpotential 
\begin{equation}
W=\frac{m_\phi}{2}\,\Phi^2+\mu^2\,\left(\beta+S\right)
\label{sup}
\end{equation}
and with minimal K\"ahler potential ${\cal {K}}=\Phi^\dagger\,
\Phi+S^\dagger\, S$. The field $\phi$ (that is, the scalar component of
$\Phi\,$) acts as the inflaton. As it is known, the corrections from the
K\"ahler potential to the scalar potential are very relevant when $\langle
\phi \rangle \gsim M_p$. This is a common problem for supersymmetric
theories of inflation, where $F$--terms usually spoil the flatness of the
inflaton potential during inflation (for a review, see \cite{lyri}). This
is in particular true for the above superpotential~(\ref{sup}), so that
additional contributions -- possibly additional scalar fields (as for
example in $D$ term inflation~\cite{dterm}) -- must be relevant during
inflation. However, we are interested in the dynamics of the system {\it
after} inflation, when  $\langle \phi \rangle \lsim M_p$, and supergravity
corrections are not important. We then assume that in the reheating stage
only one of the fields which were driving inflation is still relevant, and
we refer to it as the inflaton. We also assume $m_\phi\sim 10^{13}$ ~GeV,
as required by the COBE normalization of the CMB fluctuations for the
``usual'' chaotic inflation.

The superfield $S$ leads to the breaking of supersymmetry in the true
vacuum owing to its ``Polonyi'' superpotential~\cite{polo}. By imposing
$\beta=\left(2-\sqrt{3}\right)\,M_p$, one can indeed break supersymmetry
while retaining a vanishing cosmological constant in the true vacuum,
where the fields $s$ (the scalar component of the superfield $S$) has
expectation value of the order of $M_p$. The gravitino mass in the vacuum
is of the order $\mu^2/M_p$. In order to have a gravitino mass of about
$100$~GeV (that is the expected value for the gravitino mass in
gravity--mediated supersymmetry breaking models), $\mu \sim 10^{10}$~GeV
is required.~\footnote{In general, the presence of the Polonyi sector will
affect  the inflationary dynamics (as discussed in \cite{buc}). Again,
since we are not interested in the inflationary expansion itself, we will
not consider this issue here.} 

The superpotential $W$ describes a system where supersymmetry is broken in
a ``hidden'' sector and is transmitted gravitationally to a ``visible''
one, to whom the inflaton belongs. This separation between the two sectors is suggested by the two different mass scales which characterize inflation (inflaton mass $\sim 10^{13}$ GeV) and supersymmetry breakdown (gravitino mass $\sim 10^{2-3}$ GeV).

Right after inflation the field $\phi$ is oscillating about the bottom of its potential with frequency proportional to $m_\phi$. The time dependent expectation value of $\phi$ acts as an effective mass for the Polonyi scalar, which has vanishing expectation value at this stage. In this initial period the (time dependent) expectation value of $\phi$ is the main source of supersymmetry breaking. The amplitude of the oscillations of the field $\phi$ eventually decreases, due to the expansion of the Universe, and for times of the order of $m_{3/2}^{-1}$ the Polonyi scalar starts rolling down towards its true minimum and then oscillates about it.~\footnote{There is of course a possible moduli problem associated with these oscillations. However, we do not consider this issue here.}

The system is thus governed by two time scales. At ``early'' times, of the
order of $m_\phi^{-1}$, the only relevant dynamics is the one of the
inflaton sector, that is also the main source of supersymmetry breaking.
At ``late'' times, much larger than $m_{3/2}^{-1}$, the system behaves as
if it was in its true vacuum, and supersymmetry is broken by the Polonyi
sector.  To be more specific, we define the dimensionless parameter
$\hat{\mu}^2 \equiv \mu^2/ \left( m_\phi\,M_p \right)\sim m_{3/2}/m_\phi$,
that gives the ratio of the two time scales in the system.  If
supersymmetry is supposed to solve the hierarchy problem, $\hat{\mu}^2$
should be of the order of $10^{-11}$. Such a small parameter implies a
very large difference between the two time scales of the problem, which is
a source of technical difficulties in the numerical computations. As a
consequence, we could not study the evolution of the system for such a
small value of $\hat{\mu}^2$.  Thus, we kept it as a free parameter and we
studied how a variation of $\hat{\mu}^2$ affects the scaling of the
relevant quantities. 

\begin{figure}[h]
\centerline{\psfig{file=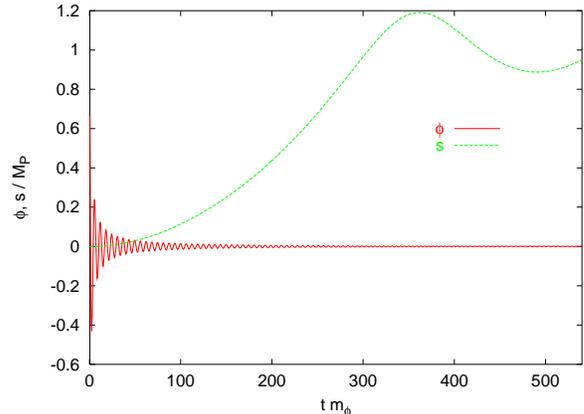,angle=-90,width=0.45 \textwidth}}
\caption{Evolution of the two scalar fields $\phi$ and $s$ for ${\hat \mu}^2 = 10^{-\,2}\,$.}
\label{fig1}
\end{figure}

To make the difference in the two time scales that govern the system manifest, we show in fig. \ref{fig1} the evolution of the inflaton and the Polonyi scalar for the case $\hat\mu^2=10^{-2}$.

We can indeed see the two different stages of the scalar evolution, the first
of which is characterized by the inflaton oscillations, while the second is
governed by the dynamics of the Polonyi field. While $\langle s \rangle$ gives
the supersymmetry breaking in the vacuum, both scalar fields contribute to
break supersymmetry during their evolution. In particular, both their kinetic
and potential energies contribute to the breaking, as emphasized
in~\cite{grt2}. Following~\cite{grt2}, we define
\begin{equation}
f_{\phi_i}^2 \equiv m_i^2 + \frac{1}{2} \, \left( \frac{d \phi_i}{d t}
\right)^2 \;\;,
\end{equation}
with $m_i = {\rm exp} \left( {\cal {K}} \, M_p^{-\,2} / 2 \right) \, \left[
\partial_i W + M_p^{-\,2} \, \partial_i \, {\cal {K}} \, W \right]\,$. The
quantities $f_i$ give a ``measure'' of the size of the supersymmetry
breaking provided by the $F$  term associated with the $i-$th scalar
field. More precisely, we will be interested in the normalized quantities
$r_i \equiv f_i^2 / \left( f_1^2 + f_2^2 \right)\,$, which indicate the
relative contribution of the two scalars.

\begin{figure}[h]
\centerline{\psfig{file=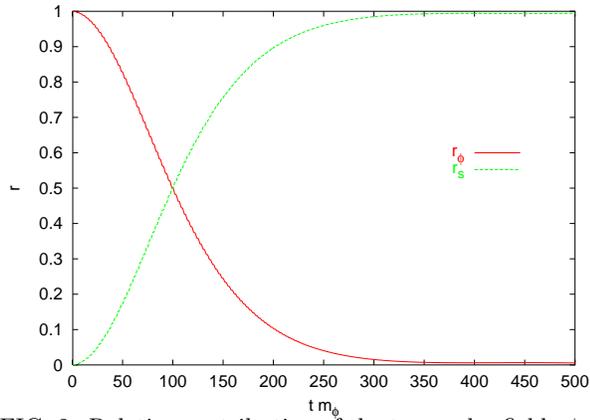,angle=-90,width=0.45 \textwidth}}
\caption{Relative contribution of the two scalar fields $\phi$ and $s$ to
the supersymmetry breaking during their evolution.
As in fig.~\ref{fig1}, ${\hat \mu}^2 = 10^{-\,2}\,$.}
\label{fig2}
\end{figure}

In fig.~\ref{fig2} we show the evolution of $r_\phi$ and $r_s$ for
the specific case ${\hat \mu}^2 = 10^{-\,2}\,$. As it occurs for all
values of ${\hat \mu}^2 \,$, in the initial stages $r_\phi \simeq 1\,$,
while $r_s \simeq 1$ at the end. The regime of equal contribution is 
around $t\, m_\phi={\hat \mu}^{-\,2}\,$.

Let us now consider the fermionic content. We denote the fermions of the
two chiral multiplets by ${\tilde \phi}$ (the ``inflatino'') and ${\tilde
s}$ (the ``Polonyino''). One linear combination of them is the goldstino
$\upsilon\,$, while the one orthogonal to $\upsilon\,$ is denoted by
$\Upsilon\,$. Initially, $\upsilon \equiv {\tilde \phi}\,$, while
$\upsilon \equiv {\tilde s}$ at late times. In addition, we have the
gravitino field, whose longitudinal and transverse component are denoted
by $\theta$ and by $\psi_i^T\,$, respectively. The transverse component is
decoupled from the other fermion fields, and its quanta are produced only
gravitationally \cite{kklv1,grt1,mar}. It will not be considered in the
remainder of this work. The longitudinal gravitino component, which in the
super-higgs mechanism is provided by the goldstino $\upsilon\,$, is
however coupled with $\Upsilon\,$. Equations of motion for the coupled
system are given in~\cite{kklv2}. The step from these equations to the
occupation numbers of the fermions $\theta$ and $\Upsilon$ is far from
trivial and requires a significant extension of the existing formalism for
nonperturbative production in the one field case. We will present the
details in a separate publication~\cite{noi}. In the remainder of this
work we will only present our results.

\begin{figure}[h]
\centerline{\psfig{file=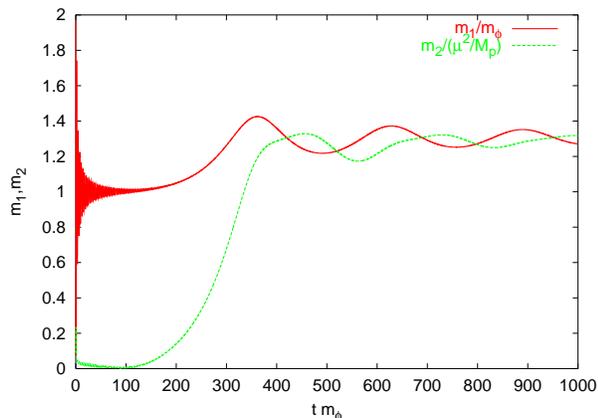,angle=-90,width=0.45 \textwidth}}
\caption{Evolution of the masses of the two fermionic eigenstates. As in
fig.~\ref{fig1}, ${\hat \mu}^2 = 10^{-\,2}\,$. Notice the different normalizations for the two masses.}
\label{fig3}
\end{figure}

As a starting point, one has to diagonalize (at each time) the coupled
$\theta$--$\Upsilon$ system. We denote the two fermionic mass eigenstates
by $\psi_1$ and $\psi_2\,$. In fig.~\ref{fig3} we show~\footnote{From
fig.~\ref{fig3}, one may be tempted to identify $\psi_1 \equiv {\tilde
\phi}$ and $\psi_2 \equiv {\tilde s}\,$. Although this identification is
rigorous only at the beginning and at the end of the evolution, it can be
used for an ``intuitive'' understanding of the system.} the evolution of
their masses for the specific case ${\hat \mu}^2 = 10^{-\,2}\,$. The most
relevant information which can be derived by this evolution is very clear:
at late times the fields $\psi_1$ and $\psi_2$ have, respectively, the
mass of the inflatino and of the gravitino field. That is, at late times
we have the identification $\psi_1 \equiv {\tilde \phi} \equiv \Upsilon
\;,\; \psi_2 \equiv \theta\;$ (${\tilde s} = \upsilon = 0 \,$, being the
goldstino). This situation is orthogonal to the initial one, when $\psi_1
\equiv \theta \;,\; \psi_2 \equiv {\tilde s} \equiv \Upsilon \;$ (${\tilde
\phi} = \upsilon = 0$). At intermediate times, when supersymmetry is
broken by both scalar fields, the gravitino is a mixture of $\psi_1$ and
$\psi_2\,$. To qualitatively appreciate the evolution of the gravitino
occupation number, we may consider $N_\theta \equiv r_1 \, N_1 + r_2 \,
N_2$ and the orthogonal combination $N_\Upsilon \equiv r_2 \, N_1 + r_1 \,
N_2\,$, where $r_i$ are the relative contributions of the two scalars to
supersymmetry breaking defined above. 

\begin{figure}[h]
\centerline{\psfig{file=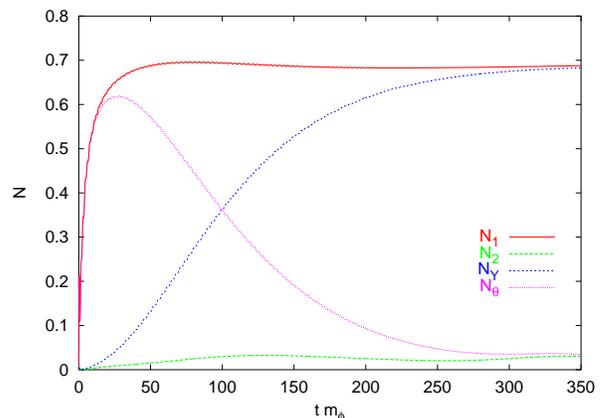,angle=-90,width=0.45 \textwidth}}
\caption{Evolution of $N_\theta$ and $N_\Upsilon$ for ${\hat \mu}^2 =
10^{-\,2}\,$ and $k = m_\phi\,$. See the text for details.}
\label{fig4}
\end{figure}

We show in fig.~\ref{fig4} the evolution of $N_1 \,,\, N_2 \,,\, N_\theta
\,,\, N_\Upsilon$ for modes of comoving momentum $k=m_\phi$ and for 
${\hat \mu}^2 = 10^{-\,2}\,$ . Notice that (by construction) $N_\theta
\equiv N_1$ at early times, while $N_\theta \equiv N_2$ at late ones. We
also see that $\psi_1$ is populated on time scales $m_\phi^{-\,1}\,$,
while $\psi_2$ on time scales ${\hat \mu}^{-\,2} \, m_\phi^{-\,1}\,$. This
feature is common for all ${\hat \mu}^2\,$~\cite{noi}. We remark that the
identification $\theta \equiv r_1 \, \psi_1 + r_2 \, \psi_2$ should be
taken only as a qualitative indication. However, the most relevant
identification $\theta \equiv \psi_2$ at late times is a rigorous one, as
should be clear from the above discussion.

\begin{figure}[h]
\centerline{\psfig{file=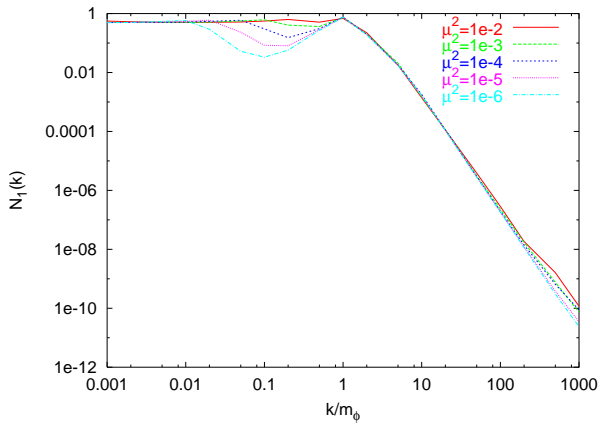,angle=-90,width=0.45 \textwidth}}
\caption{Spectrum of inflatinos at late times.}
\label{fig5}
\end{figure}

\begin{figure}[h]
\centerline{\psfig{file=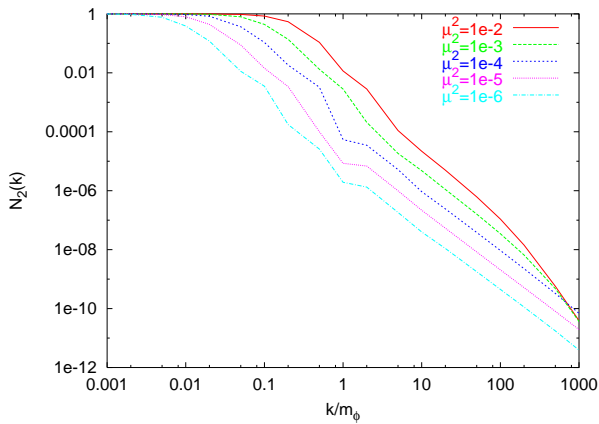,angle=-90,width=0.45 \textwidth}}
\caption{Spectrum of gravitinos at late times.}
\label{fig6}
\end{figure}

We are now ready to present our most important result: the occupation
number of $\Upsilon$ and $\theta$ at the end of the process. We show them
in figs.~\ref{fig5} and \ref{fig6}, respectively.~\footnote{These spectra
are shown at the time $t=10 \, {\hat\mu}^{-\,2} m_\phi^{-\,1}\,$. In the
${\hat \mu} = 10^{-\,2} - 10^{-\,4}$ cases we have continued the evolution
further, until the spectra stop evolving. We have found that the spectra
shown in fig.~\ref{fig5} coincide with the final ones, while $N_\theta$
very slightly {\em decreases} for $t > 10 \, {\hat\mu}^{-\,2}
m_\phi^{-\,1}\,$. Thus, the results shown in fig.~\ref{fig6} give an
accurate upper bound on the final gravitino abundance.}

The time required for the numerical computation increases linearly with
${\hat \mu}^2\,$, so that we could perform it up to ${\hat \mu} =
10^{-\,6}\,$. The realistic case ${\hat \mu}^2 = 10^{-\,11}$ is far from
our available resources, so we kept $\hat\mu^2$ as a free parameter. The
case ${\hat \mu}^2 = 10^{-\,11}$ can be clearly extrapolated from the
results we are going to discuss. Moreover, the case $\hat{\mu}^2=0$ can be
studied analytically, and it agrees with the limit ${\hat \mu} \rightarrow
0$ that one deduces from the numerical results. For ${\hat \mu} = 0\,$,
only the inflatino is produced. The mass of the Polonyi fermion does
instead vanish identically, so no quanta of this particle are produced at
preheating~\cite{noi}. Notice that ${\hat \mu} = 0$ corresponds to a
situation with unbroken susy in the vacuum, and it reproduces the models
with one single field studied so far. We stress that in this
case only the inflatino is produced at preheating.

In fig.~\ref{fig5} the spectrum of $\psi_1$ in shown for
$\hat{\mu}^2=10^{-2},\,10^{-3},\,10^{-4},\,10^{-5},\,10^{-6}$. It is apparent
that the main features of the spectrum are independent of the value of
$\hat\mu^{-2}$. The reason for this is that this eigenstate is associated to
the inflatino, that is produced by the coherent oscillations of the inflaton.
The inflaton dynamics occurs on time scales of the order of $m_\phi^{-1}$, and
is independent on $\hat\mu^2$. The spectrum given in fig.~\ref{fig6} shows a more significant dependence on $\hat\mu^2$. In particular, as in the case of the spectrum in fig.~\ref{fig5}, the occupation number is of the order of unity if the comoving momenta are smaller than some cut--off $k_*$. At variance with the spectrum of $\psi_1\,$, however, here the cut--off depends on the value of $\hat\mu^2$. The total number of particles produced in this case is a increasing function of $\hat\mu^2$. 

It is possible to estimate in a very simple way how $k_*$ scales with $\hat\mu^2$. The cut--off in the spectrum (in terms of the comoving momentum) of particles produced at preheating is indeed generically proportional to $\sqrt{\dot{m}}\,a$, where $\dot{m}$ is the time--derivative of the effective mass of the produced particle, and $a$ is the scale factor of the Universe at the time of preheating. In the case of the Polonyino, the typical mass scale is $m_{3/2}$, while the typical time scale is $m_{3/2}^{-1}$. As a consequence, $\sqrt{\dot{m}}\sim m_{3/2}$. When the energy in the Universe is dominated by the sinusoidal oscillations of a massive scalar field, $a\left( t\right)\sim t^{2/3} \sim \left(m_{3/2}\right)^{-2/3}$. Collecting all these estimates, we obtain
$k_*\sim m_\phi\,\left(\hat\mu^2\right)^{1/3}\;.$
Although this is quite a simple estimate, it shows very good agreement with the numerical results. Moreover, it agrees with the fact that $N_2=0$ for $\hat\mu^2=0$.

In conclusion, our calculation confirms that one fermionic eigenstate is
efficiently produced at preheating. However, as we have initially
remarked, in more realistic models (i.e. with more than one chiral
superfield and supersymmetry broken in the vacuum) this fermionic field
should be regarded as the ``inflatino'' rather than the ``gravitino''. We
have shown that gravitino production is significantly reduced if the sector
responsible for supersymmetry breaking today is coupled only
gravitationally to the one responsible for inflation. The reason for this
small production is related to the smallness of the supersymmetry breaking
scale (i.e. of ${\hat \mu}^2$) with respect to the scale of inflation.

\acknowledgments
We are pleased to thank R.~Kallosh, L.~Kofman, A.~Linde, D.H.~Lyth, A.~Riotto, and A.~Van Proeyen for interesting conversations. This work is supported by the European Commission RTN programmes HPRN-CT-2000-00148 and 00152.


\begin{thebibliography}{99}
\frenchspacing 


\bibitem{tm}
T.~Moroi, Ph-D thesis, Tohoku, Japan, hep-ph/9503210.

\bibitem{gprt}
G.~F.~Giudice, M.~Peloso, A.~Riotto and I.~Tkachev, JHEP{\bf 9908}, 014 (1999).

\bibitem{kklv1}
R.~Kallosh, L.~Kofman, A.~Linde and A.~Van Proeyen,
Phys.\ Rev.\ {\bf D 61}, 103503 (2000).

\bibitem{grt1}
G.~F.~Giudice, I.~Tkachev and A.~Riotto, JHEP {\bf 9908}, 009 (1999).

\bibitem{grt2}
G.~F.~Giudice, A.~Riotto and I.~Tkachev, JHEP {\bf 9911}, 036 (1999).

\bibitem{kklv2} 
R.~Kallosh, L.~Kofman, A.~Linde and A.~Van Proeyen, Class.\ Quant.\ Grav.\ {\bf 17}, 4269 (2000).

\bibitem{kim}
D.~H.~Lyth and H.~B.~Kim, hep-ph/0011262.

\bibitem{lyri}
D.~H.~Lyth and A.~Riotto, Phys.\ Rept.\ {\bf 314}, 1 (1999).

\bibitem{dterm}
P.~Binetruy and G.~Dvali, Phys.\ Lett.\ B {\bf 388}, 241 (1996); E.~Halyo, Phys.\ Lett.\ B {\bf 387}, 43 (1996).

\bibitem{polo}
J.~Polonyi, {\it  Hungary Central Inst Res - KFKI-77-93 (77,REC.JUL 78) 5p}.

\bibitem{buc}
W.~Buchmuller, L.~Covi and D.~Delepine, Phys.\ Lett.\ B {\bf 491}, 183 (2000).

\bibitem{mar}
A.~L.~Maroto and A.~Mazumdar, Phys.\ Rev.\ Lett.\ {\bf 84}, 1655 (2000).

\bibitem{noi}
H.P.~Nilles, M.~Peloso, and L.~Sorbo, to appear.



\end{thebibliography}
\end{document}